\makeatletter
\let\old@footnotetext=\@footnotetext
\makeatother
\documentclass{aastex}
\makeatletter
\let\@footnotetext=\old@footnotetext
\makeatother
\usepackage{emulateapj5}
\shorttitle{Iron-Rich Ejecta in the LMC SNR DEM L71}
\shortauthors{Hughes et al.}
\submitted{Received 2002 November 20; accepted 2002 December 4}
\accepted{To appear in the Astrophysical Journal Letters 2003 January 10}
\def\ha{H$\alpha$}

\def\chandra{{\it Chandra}}
\def\dem71{DEM~L71}
\def\einstein{{\it Einstein}}

\def\asca{{\it ASCA}}
\def\astroe2{{\it Astro-E2}}
\def\xmm{{\it XMM-Newton}}
\def\chandra{{\it Chandra}}

\def\lsim{\hbox{\raise.35ex\rlap{$<$}\lower.6ex\hbox{$\sim$}\ }}
\def\gsim{\hbox{\raise.35ex\rlap{$>$}\lower.6ex\hbox{$\sim$}\ }}

\begin{document}

\title{Iron-Rich Ejecta in the Supernova Remnant DEM L71}

\author{John P.~Hughes\altaffilmark{1},
Parviz Ghavamian\altaffilmark{1},
Cara E.~Rakowski\altaffilmark{1}, and
Patrick O.~Slane \altaffilmark{2},
}
\altaffiltext{1}{Department of Physics and Astronomy, Rutgers
University, 136 Frelinghuysen Road, Piscataway, NJ 08854-8019}
\altaffiltext{2}{Harvard-Smithsonian Center for Astrophysics, 60 Garden Street,
Cambridge, MA 02138}

\begin{abstract}

\chandra\ X-ray observations of \dem71, a supernova remnant (SNR) in
the Large Magellanic Cloud (LMC), reveal a clear double shock
morphology consisting of an outer blast wave shock surrounding a
central bright region of reverse-shock heated ejecta.  The abundances
of the outer shock are consistent with LMC values, while the ejecta
region shows enhanced abundances of Si, Fe, and other species.
However, oxygen is not enhanced in the ejecta; the Fe/O abundance
ratio there is $\gsim$5 times the solar ratio.  Based on the relative
positions of the blast wave shock and the contact discontinuity in the
context of SNR evolutionary models, we determine a total ejecta mass
of $\sim$1.5 $M_\odot$. Ejecta mass estimates based on emission
measures derived from spectral fits are subject to considerable
uncertainty due to lack of knowledge of the true contribution of
hydrogen continuum emission. Maximal mass estimates, i.e., assuming no
hydrogen, result in 1.5 $M_\odot$ of Fe and 0.24 $M_\odot$ of
Si. Under the assumption that an equal quantity of hydrogen has been
mixed into the ejecta, we estimate 0.8 $M_\odot$ of Fe and 0.12
$M_\odot$ of Si.  These characteristics support the view that in
\dem71\ we see Fe-rich ejecta from a Type Ia SN several thousand years
after explosion.

\end{abstract}


\section{Introduction}

A distinguishing characteristic of Type Ia supernovae (SNe) is the
large amount of iron that they are thought to produce mainly through
the decay of radioactive Ni$^{56}$. The nuclear decay of this material
powers the SN light curve, thereby providing a constraint on the
quantity of material ejected in individual SN events. Although
variation in the amount and distribution of radioactive ejecta likely
accounts for the modest range of peak absolute magnitudes seen in Type
Ia SNe, by and large SNe Ia are strikingly homogeneous in their
properties. Various observational constraints (e.g., the absence of
hydrogen in optical spectra near optical maximum, occurrence in older
stellar populations, and the relatively large amount of Ni$^{56}$
produced) have led to the current framework involving carbon
deflagration/detonation in a white dwarf driven to the Chandrasekhar
limit by accretion. The most likely progenitor system (Branch et
al.~1995) is a C-O white dwarf accreting H/He-rich gas from a
companion, either from its wind or through Roche lobe overflow.
Remnants of SN~Ia should contain about one Chandrasekhar mass of
ejecta comprised of 0.6--0.8 $M_\odot$ of iron with intermediate-mass
elements (O, Mg, Si, S, Ca) distributed in the outer layers (e.g.,
Iwamoto et al.~1999).  Current models predict that SNe Ia should leave
no compact remnant.

One of the goals of supernova remnant (SNR) research has been to use
remnant properties to infer the nature of the progenitor. Such
properties as the partially neutral nature of the surrounding ambient
medium (Tuohy et al.~1982) or the Fe-rich composition of the ejecta
(Hughes et al.~1995) have been used to argue for a SN~Ia origin for
some SNRs.  In this article we investigate the Large Magellanic Cloud
(LMC) SNR \dem71\ (0505$-$67.9) which exhibits both of these features.
\dem71\ (Davies, Elliott, \& Meaburn 1976) was first proposed as a
supernova remnant based on the \einstein\ X-ray survey of the LMC
(Long, Helfand, \& Grabelsky 1981).  Follow-up optical spectroscopy
confirmed this identification and revealed the remnant to be
Balmer-dominated (Tuohy et al.~1982; Smith et al.~1991), that is to
say, one whose optical spectrum is dominated by hydrogen emission with
little or no emission from forbidden lines of [\ion{O}{3}] or
[\ion{S}{2}].  The remnants of Tycho's SN and SN1006 are Galactic
examples of such SNRs (Kirshner \& Chevalier 1978; Schweizer \& Lasker
1978).  \asca\ X-ray studies of \dem71\ (Hughes, Hayashi, \& Koyama
1998) yielded an age of $\sim$5000 yr and evidence for the presence of
SN ejecta in the form of an enhanced global abundance of Fe.  Here we
present results from new \chandra\ observations of the remnant that
bear closely on the nature of the originating SN.

\section{Analysis}

We observed \dem71\ using the back-side-illuminated chip (S3) of the
Advanced CCD Imaging Spectrometer (ACIS-S) instrument in full-frame
timed exposure mode starting on 2000 January 04 for 45.4 ks (OBSID
775). The reduction of these data proceeded as described in Rakowski
et al.~(2002). Our \ha\ image was constructed from a subset of a 3-D
position-velocity data cube obtained on \dem71\ using the Rutgers
Fabry-Perot (RFP) at the focus of the CTIO 1.5-m on 1998 February 18.
The RFP data reduction is described in Ghavamian et al.~(2002).

\begin{figure*}[t]
\vspace{-5truein}
\epsfbox{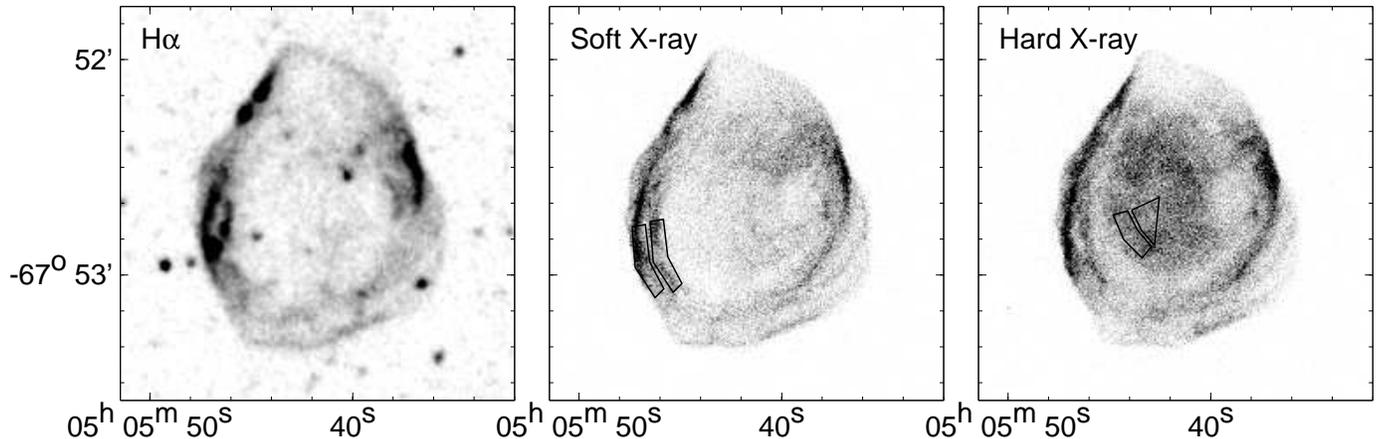}
\vspace{10pt}
\figcaption {Images of the supernova remnant \dem71\ in \ha\
(left panel); {\em Chandra} soft X-ray (0.2--0.7 keV) (middle panel);
and {\em Chandra} hard X-ray band (0.7--2.6 keV) (right panel).  North
is up and east is to the left; axes are labelled in epoch J2000. The
grayscale intensity scaling is linear in all panels. Spectral
extraction regions are superposed on the soft and hard X-ray images.
}
\end{figure*}

\subsection{Spatial}

The soft X-ray image of \dem71\ from \chandra\ (Fig.~1, middle panel)
shows emission from an egg-shaped outer rim that matches nearly
perfectly the optical \ha\ image (Fig.~1, left panel). Features in
common include matched variations in surface brightness around the rim
as well as the presence of multiple nested filaments, which are
particularly noticeable along the southern edge.  Faint diffuse
emission in the interior of the SNR agrees well too. The remarkable
detailed similarity of the X-ray and optical morphology is uncommon
among supernova remnants in general.  Nearly all of the optical
emission in \dem71\ arises in high velocity (500--1000 km s$^{-1}$)
nonradiative shocks, with little contribution from the slow ($\sim$100
km s$^{-1}$) radiative shocks typically seen in middle-aged remnants.
Furthermore the X-ray emission shows broadly similar intrinsic
emissivities around the rim (there are no large variations in
abundance or thermodynamic state, see Rakowski, Ghavamian, \& Hughes
2002) so that surface brightness variations largely trace variations
in the ambient gas density around the remnant.  North-south the outer
rim spans 83$^{\prime\prime}$, while in the east-west direction it is
only 60$^{\prime\prime}$ across.  Any differences between the X-ray
and optical sizes are small, no more than a 0.5$^{\prime\prime}$
difference in diameter, with the optical dimensions being marginally
larger than the X-ray ones.

The hard X-ray image of \dem71\ (Fig.~1, right panel) shows emission
from the outer rim that resembles the soft X-ray and \ha\ images.
However, this band also reveals a new spatial component of X-ray
emission that has no counterpart in the other bands.  This emission
almost fills the interior and appears as a modestly limb-brightened,
elliptical shell of emission slightly offset from the geometric center
of the outer rim. The emission extends 40$^{\prime\prime}$ by
32$^{\prime\prime}$ at a position angle (for the major axis) of
$\sim$12$^\circ$ east of north and is centered at 05$^{\rm h}$05$^{\rm
m}$42.6$^{\rm s}$ $-$67$^{\circ}$52$^{\prime}$38.6$^{\prime\prime}$
(J2000).  Hereafter we refer to this as the ``core'' emission.

\subsection{Spectral}

Spectra were extracted from four regions in the remnant's southeastern
quadrant (see figure 1) to investigate the soft and hard X-ray
emission components.  Two spectra of the soft rim (``outer'' and
``inner''), corresponding to the two nested filaments that appear
here, were extracted.  Likewise two spectra of the core X-ray emission
(again ``outer'' and ``inner'') were extracted.  Background-subtracted
count rates are given in Table 1. Background represented less than
0.5\% of the source count rate in the 0.2--5 keV band as taken from a
1$^\prime$--2$^\prime$ annulus outside the remnant. Before fitting,
source spectra were rebinned to a minimum of 10 counts per channel.
The rim data sets (2 left panels of fig.~2) are quite similar as are
the core data sets (2 right panels), but the rim and core spectra
differ significantly from each other.

\begin{figure*}[t]
\begin{center}
\epsfxsize=18truecm\epsfbox{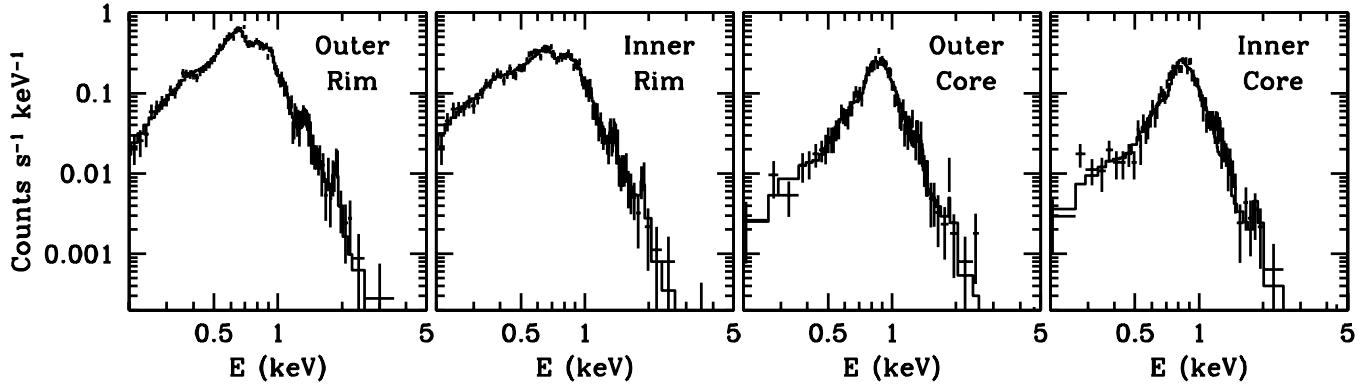}
\end{center}
\vspace{-10pt}
\figcaption {X-ray spectra extracted from several locations
in the southeastern quadrant of \dem71\ with best-fit nonequilibrium
ionization thermal plasma emission models.
}
\end{figure*}

The abundances and emission properties of the gas were constrained
through fits of nonequilibrium ionization (NEI) planar-shock thermal
plasma models (see Hughes, Rakowski, \& Decourchelle
2000). The abundances of O, Ne, Mg, Si, S, and Fe were allowed
to vary, while C, N, Ar and Ca were fixed at 20\% of
solar and the Ni abundance was tied to Fe. The fits were
generally acceptabl weith similar best-fit values for similar regions
(see Table 1).  Absorption of low energy X-rays by intervening gas and
dust was included with the hydrogen column density, $N_{\rm H}$, as a
free parameter.  The values obtained, $N_{\rm H}$ = 4--$7 \times
10^{20}$ atoms cm$^{-2}$, are consistent with previous X-ray spectral
fits (Hughes et al.~1998) and the Galactic \ion{H}{1} column density
in this direction (Dickey \& Lockman 1999).

{\parindent=0cm 
\begin{tiny}
\begin{minipage}[t]{0.47\textwidth}
\begin{center}
{\noindent{TABLE 1}}\\
{\noindent{\sc NEI Thermal Model Fits for Regions in \dem71}}\\[6pt]
\begin{tabular}{@{}lcccc@{}}
\hline\hline\\[-4pt]
Parameter & Outer rim & Inner rim & Outer core & Inner core\\[4pt]
\hline\\[-4pt]
Rate (s$^{-1}$) & 0.29 & 0.20 & 0.098 & 0.093 \\[2pt]
$N_{\rm H}$ ($10^{21}$ cm$^{-2}$) & $0.58^{+0.04}_{-0.06}$ & $0.41^{+0.04}_{-0.04}$
 & $0.67^{+0.31}_{-0.17}$  & $0.52^{+0.24}_{-0.14}$ \\[2pt]
$kT$ (keV)                   & $0.47^{+0.03}_{-0.03}$ & $0.47^{+0.03}_{-0.04}$
 & $1.6^{+0.7}_{-0.5}$ & $1.1^{+0.3}_{-0.3}$ \\[2pt]
$\log(n_et/{\rm cm^{-3} s})$ & $11.64^{+0.11}_{-0.11}$& $11.66^{+0.15}_{-0.20}$
 & $10.80^{+0.22}_{-0.16}$ & $11.00^{+0.31}_{-0.16}$\\[2pt]
O                            & $0.21^{+0.03}_{-0.02}$ & $0.13^{+0.02}_{-0.01}$
 & $0.31^{+0.25}_{-0.09}$ & $0.33^{+0.27}_{-0.09}$ \\[2pt]
Ne                           & $0.42^{+0.05}_{-0.04}$ & $0.34^{+0.04}_{-0.03}$
 & $2.3^{+1.9}_{-0.9}$ & $1.7^{+1.5}_{-0.8}$ \\[2pt]
Mg                           & $0.37^{+0.08}_{-0.07}$ & $0.24^{+0.07}_{-0.07}$
 & $1.7^{+1.6}_{-0.7}$ & $1.1^{+1.2}_{-0.5}$ \\[2pt]
Si                           & $0.33^{+0.11}_{-0.11}$ & $0.23^{+0.14}_{-0.12}$
 & $0.7^{+1.0}_{-0.5}$ & $1.0^{+1.1}_{-0.5}$ \\[2pt]
S                            & $0.06^{+0.30}_{-0.06}$ & $<$0.30
 & $<$2.6 & $<$2.1 \\[2pt]
Fe                        & $0.093^{+0.012}_{-0.009}$ & $0.079^{+0.010}_{-0.009}
$
 & $1.9^{+1.4}_{-0.6}$ & $1.7^{+1.2}_{-0.6}$ \\[2pt]
$\chi^2$/d.o.f               &  103.1/100  & 79.4/91 & 85.6/73 & 102.6/69 \\[4pt
]
\hline\\[-4pt]
\end{tabular}
\end{center}
\end{minipage}
\end{tiny}
}

\par
The ionization timescale of the rim region indicates that the gas is
out of equilibrium, as expected.  
The fitted abundances are sub-solar and generally consistent with the
gas phase LMC values (Russell \& Dopita 1992). The Fe abundance is
derived from the L-shell lines that produce, at CCD spectral
resolution, a quasi-continuum blend of emission over the 0.7--1.5 keV
band. Note that our NEI spectral model is derived from the original
Raymond \& Smith (1977) code (and subsequent revisions) which accurately 
models the overall flux
of the Fe L-shell emission, but is less accurate at modeling its
detailed spectral variation with photon energy, due to limitations in
the availability of atomic data.  This is the case to a greater or
lesser extent for all NEI spectral models currently
available. Typically with such models, the fitted Fe abundance is low,
while the abundances of Ne and Mg, whose K-shell lines fall in this
energy band, are enhanced to compensate for deficiencies in the
modeled Fe L-shell emission.  Nevertheless these spectral fits (as
well as the more comprehensive study of \dem71's outer rim in Rakowski
et al.~2002) confidently confirm that the rim region is swept-up
interstellar medium (ISM) heated to X-ray emitting temperatures by a
high velocity blast wave.  We note that there is no evidence for a
hard power-law tail beyond the thermal emission in the \chandra\
spectrum of the rim.

In marked contrast to the rim regions, the core spectra are dominated
by the Fe L-shell blend, which peaks near 0.8 keV, and have no
significant O line emission at $\sim$0.65 keV.  The fitted parameters
(Table 1) reveal that the abundances of all species, with the
exception of O, are considerably greater than the LMC mean, while the
abundances of some species, Ne and Fe in particular, are even greater
than solar.  The abundance ratio of iron to oxygen is $\gsim$5 times
the solar ratio.  These large absolute abundances, in addition to the
non-solar abundance ratios, signal the presence of supernova ejecta in
the core of the remnant.

\section{Ejecta Mass}

Here we quantitatively examine the properties of the core region to
identify the evolutionary state of the remnant and the nature of the
originating SN.  The large iron-to-oxygen ratio of the core X-ray
emission, as well as its highly-symmetric, well-ordered spatial
distribution, immediately suggests a SN~Ia origin for \dem71.  A
critical diagnostic of the SN type is the total mass of ejecta.  We
estimate this in two ways, first using dynamical considerations and
then from the properties of the core X-ray emission.

The position of the reverse shock (or contact discontinuity) relative
to that of the blast wave can constrain the evolutionary state of a
remnant in the context of self-similar models (e.g., Truelove \& McKee
1999). Once the profiles of the ISM and ejecta are chosen, this
constraint is unique. Although subject to hydrodynamic instabilities,
the position of the contact discontinuity between the SN ejecta and
the ambient medium is reasonably well identified in our image --- it
is marked by the outer extent of the core hard emission. This has a
geometric mean radius of 4.3 pc assuming an LMC distance of 50 kpc.
\dem71's outer blast wave has a geometric mean radius of 8.6 pc,
yielding a ratio for the radii of $\sim$2.

In their study of instabilities and clumping in SNe Ia, Wang \&
Chevalier (2001) present the 1-D radial evolution of the contact
discontinuity and blast wave for an exponential ejecta density profile
interacting with a constant density ambient medium. For this model,
the radius of the contact discontinuity is approximately one-half that
of the blast wave when the normalized radius of the forward shock is
$r^\prime \sim 3$. The physical radius of the forward shock, $r$,
depends on the normalized radius, $r^\prime$; the ejected mass, $M$;
and the ambient hydrogen number density, $n_0$, through a scaling
relation.  Inverting, we obtain a relation for the ejected mass in
terms of the Chandrasekhar mass, $M_{\rm ch}$
$$M = M_{\rm ch}\, n_0\, 
\left ( {r/1~{\rm pc}\over 2.19\, r^\prime} \right )^3 .$$
\noindent
which evaluates to $M \sim 1.1 M_{\rm ch}$ for $r^\prime = 3$,
$r=8.6$ pc and a mean ambient density of 0.5 cm$^{-3}$ 
%
%
(Ghavamian et al.~2002).  At this time the entire ejecta
have been thermalized, the reverse shock has just reached the center, 
and the ejecta are moving outward relatively slowly at $\lsim$
100 km s$^{-1}$.

For our second ejecta mass estimate, we extracted spatially resolved
spectra of the core from five concentric elliptical rings spaced by
5$^{\prime\prime}$ (in semi-major axis length).  The annuli had
constant ellipticity (an axial ratio of 1.25) and position angle
(12$^\circ$ east of north).  Each of these spectra contain projected
emission from the blast wave, which we removed by subtracting off a
spectrum of the bright rim emission scaled independently for each
annular ring to zero out the resultant spectrum over the 0.2--0.6
keV band. This band was chosen because it shows no evidence for
enhanced central emission and thus plausibly corresponds to emission
from the blast wave alone.  The count rates of the various spectra
range from 0.068 s$^{-1}$ to 0.38 s$^{-1}$.

These spectra all show the broad Fe L-shell blend and clearly resolved
Si and S K$\alpha$ lines, but no obvious K$\alpha$ lines of O, Ne, or
Mg. We fit these spectra with the NEI spectral model including a
minimal set of elemental species: Si, S, Ar, Ca, Fe, Ni, and a H
continuum. To further simplify the model we restricted the number of
free parameters in the fits.  The global NEI parameters, temperature
and ionization timescale, and the Ni to Fe ratio were constrained to
be the same for all five spectra. We allowed the emission measures
(EMs) of Si and Fe to vary freely in all the spectra; however, the S,
Ar, and Ca EMs varied along with Si at their relative solar abundance
ratios. EMs are defined as $n_e n_i V$, where $n_e$ and $n_i$ are the
electron and ion number densities and $V$ is the emitting volume.
We fit both the low ($E<1.5$ keV) and high ($E>1.5$ keV) energy
portions of the spectra separately in order to ensure that the fits
were not being dominated by the ``shape'' of the Fe L-shell blend.  In
fact both portions of the spectra are broadly consistent with $kT \sim
1$ keV and $\log(n_et/{\rm cm^{-3}\,}) \sim 11.2$. Finally the fits
were performed for three values of the column density, $N_{\rm H} =
(4, 8, 12) \times 10^{20}$ atoms cm$^{-3}$, fixed separately for three
different sets of fits, in order to estimate uncertainties on the
parameters.  The fits are not particularly good from a $\chi^2$
point-of-view, but most of the residuals are associated with the Fe-L
shell blend.

From the fits we obtain numerical values for the EMs of
the Fe and Si-group elements as a function of radius in the core
region.  For simplicity we assume spherical geometry for the
deprojection.  If the true 3-D shape of the core region were
ellipsoidal, then our mass estimates would be some 13\% higher
(prolate) or lower (oblate).  We assume that the Fe and Si-group
components are co-spatial, although we allow for the components to
contribute in differing proportions in each radial shell.  The mean
charge state of each species in the fit is calculated using the values
quoted above for $kT$ and $n_et$.  For example, the mean number of
electrons per ion is 11.8 for Si and 18.2 for Fe.

\begin{center}
\begin{minipage}[t]{0.47\textwidth} 
\epsfxsize=0.98\textwidth \epsfbox{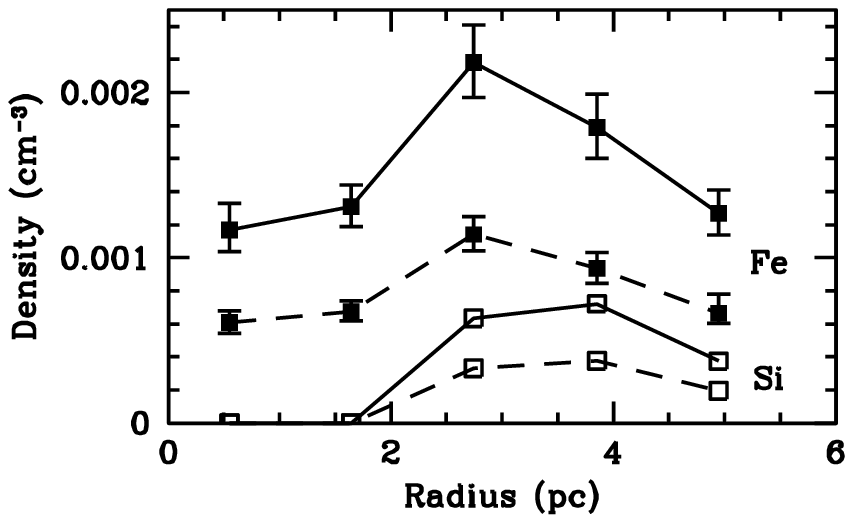}
\figcaption{Deprojected Fe and Si number densities as a
function of radius in the ejecta of \dem71. The curves with solid
(open) symbols correspond to the Fe (Si) density.  Solid curves are
the maximal density estimates; dashed curves assume an equal mass of
hydrogen mixed into the ejecta (see text). Error bars show the effect
of varying the hydrogen column density (not plotted for Si, since
these errors are smaller than the size of the points).
}
\end{minipage}
\end{center}

The main uncertainty in converting EMs to Fe and Si density and mass
estimates is not knowing which species contribute electrons.  In a
solar abundance plasma, there are $\sim$$3\times10^4$ electrons
(mostly from hydrogen) per Fe ion, compared to the roughly 50
electrons per Fe ion in a pure metal plasma. Thus if even just a small
quantity of hydrogen were mixed into the ejecta, it would drastically
affect the derived densities.  Our spectral fits are inconclusive
regarding the amount of hydrogen present in the core region.  The
spectral fits discussed here require a relatively high level of
broadband continuum emission, yet in other fits we performed a wide
range of continuum intensities were inferred, while the EMs of the
metals remained fairly stable.  Furthermore, although we have modeled
a hydrogen continuum, any fully stripped ion could produce the
requisite featureless spectrum.  Thus in lieu of a definitive result
we present a couple of astrophysically plausible estimates.  For one
we assume that the metals are the sole source of electrons, which
yields maximal density estimates.  For the other we make the plausible
assumption that a comparable mass of hydrogen has been well mixed into
the metal rich ejecta during its evolution.

The deprojected Fe and Si densities versus radius are plotted in
figure 3.  Each clearly shows enhanced density around 3--4 pc
indicating a thick shell structure with the Si slightly ahead of the
Fe.  Fe is present even in the center of the remnant, while Si is
absent there.  Integrating the density profiles we obtain total masses
of 1.5 $M_\odot$ of Fe and 0.24 $M_\odot$ of Si assuming a pure metal
composition for the emitting material. The curves for the case with a
modest admixture of hydrogen are quite similar in shape to those of
the pure metal case (see figure 3), although the ion densities we
infer are about a factor of two lower. The total amount of Fe is 0.8
$M_\odot$ and that of Si is 0.12 $M_\odot$.

\section{Summary}

The spatial and spectral analysis of the \chandra\ ACIS data reveal
\dem71\ to be a textbook example of a supernova remnant: an outer
blast wave shock propagating through the ISM that surrounds a core of
SN ejecta heated to X-ray--emitting temperatures by the inward moving
reverse shock. Our analysis strongly argues for a Type Ia SN origin
for \dem71.  Specifically we find (1) a large Fe to O ratio in the
ejecta X-ray emission; (2) $\sim$1.5 $M_\odot$ of ejecta from
dynamical considerations; (3) a relatively low mass ratio of Si to Fe
($\sim$0.15); and (4) a large mass of ejected Fe ($\sim$0.8
$M_\odot$). These results are all consistent with SN~Ia ejecta and
further may have some power to discriminate between models for how the
flame front propagates in these explosions (see, e.g., Iwamoto et
al.~1999).  The ejecta are mostly contained within a thick shell and
are partially stratified. According to our deprojection analysis, Fe
extends throughout the entire ejecta even into the center, while Si
appears to be contained only within the outer half of the ejecta.

A considerably deeper \chandra\ observation of \dem71\ has been
approved for observation in cycle 4.  This will allow us to search for
and study spectral variations within the Fe-rich ejecta and to
determine its morphology in greater detail. Measuring the kinematics
of the Fe-rich ejecta is an important next step. Unfortunately this is
likely to be beyond the capabilities of \chandra, \xmm, and even
\astroe2, although the Constellation-X Facility, if it attains angular
resolution of 10$^{\prime\prime}$, should be able to detect the
expected motions of $\sim$100 km s$^{-1}$. We are also pursuing
ground-based optical searches for coronal [\ion{Fe}{14}] $\lambda$5303
\AA\ line emission from the ejecta, which could provide complementary
insights into the nature and dynamics of the ejecta.

\acknowledgments

We thank Ted Williams for assistance with the Fabry-Perot
observations, Leisa Townsley for allowing us to use her CTI corrector
software, and John Nousek and Dave Burrows for help with the original
proposal.  Partial financial support was provided by NASA contract
NAS8-39073 (SAO), \chandra\ grants GO0-1035X and GO1-2052X (Rutgers),
and a NASA Graduate Student Researchers Program fellowship to CER.

\end{document}